\documentclass[aps,pre,showpacs,amsmath,amssymb,amsfonts,superscriptaddress,twocolumn,lengthcheck]{revtex4-1}

\usepackage{graphicx}
\usepackage{subfigure}
\usepackage{verbatim}
\usepackage{dcolumn}% Align table columns on decimal point
\usepackage{bm}% bold math
\usepackage{epsf}
\usepackage{color}
\usepackage[colorlinks=true,citecolor=blue,linkcolor=blue,urlcolor=blue]{hyperref}
\usepackage{hhline}

\usepackage[normalem]{ulem}

\newcommand{\bla}{bla\\bla\\bla\\bla\\bla}

\begin{document}

\title{Microcanonical Szil\'ard engines beyond the quasistatic regime}

\author{Thiago V. Acconcia} 
\email[]{thiagova@ifi.unicamp.br}
\affiliation{Instituto de F\'\i sica `Gleb Wataghin', Universidade Estadual de Campinas, Rua S\'ergio Buarque de Holanda, 777,  13083-859 Campinas, SP, Brazil}
  
\author{Marcus V. S. Bonan\c{c}a}
\email[]{mbonanca@ifi.unicamp.br}
\affiliation{Instituto de F\'\i sica `Gleb Wataghin', Universidade Estadual de Campinas, Rua S\'ergio Buarque de Holanda, 777,  13083-859 Campinas, SP, Brazil} 


\date{\today}

\begin{abstract} 
We discuss the possibility of extracting energy from a single thermal bath using microcanonical Szil\'ard engines operating in finite time. This extends previous works on the topic which are restricted to the quasistatic regime. The feedback protocol is implemented based on linear response predictions of the excess work. It is claimed that the underlying mechanism leading to energy extraction does not violate Liouville's theorem and preserves ergodicity throughout the cycle. We illustrate our results with several examples including an exactly solvable model.
\end{abstract}

\maketitle

\section{Introduction}

Over the last twenty years there has been an intense research activity on the interplay between information and thermodynamics. This field of investigation dates back to Maxwell who introduced a \textit{gedanken} experiment where a being, the so-called Maxwell's demon, is able to violate the second law based on the knowledge acquired about the microscopic states of the system of interest \cite{Maxwell1,Maxwell2,Vedral2009}. This paradox was later on reformulated by Szil\'ard \cite{Szilard1929}, who devised an engine able to extract energy from a single thermal bath. In Szil\'ard's setup, not only the information gathered by the demon is more easily quantified but also its role on the conversion of heat into work becomes more transparent. Hence it has become crucial to understand how thermodynamic entropy and information are related. 

Major efforts on this direction have been made, for instance, by Brillouin, Landauer, Bennett and Penrose \cite{Brillouin1951,Landauer1991,Bennett1982,Penrose1970} who have discussed the energetic costs of measurement and measurement erasure. Along these lines, the demon operating the Szil\'ard engine needs a memory device to store the information gathered by him. To reset this memory at the end of the cyclic process, the demon has to dissipate an amount of energy that is generally greater than the extracted work. Hence the second law is rescued within this scenario by attributing an entropic cost for the measurement erasure. Besides the conceptual debate, several experimental setups have recently analysed these issues \cite{Toyabe2010,Berut2011,Koski2014,Roldan2014}. 

In the last decade, feedback controlled processes such as those performed by a Maxwell's demon have been incorporated into the frameworks of fluctuation theorems \cite{Sagawa2008,Sagawa2009,Sagawa2010,Horowitz2010,Ponmurugan2010,Sagawa2012a,Sagawa2012b} and stochastic thermodynamics \cite{Esposito2011,Abreu2011,Strasberg2013,Barato2014a,Barato2014b} leading to significant progress on the history of information in thermodynamics. More recently, the demon itself has been modeled as a physical system, yielding an antonomous formulation of the original paradox \cite{Mandal2012,Barato2013,Deffner2013a,Deffner2013b,Lu2014,Pekola2016,Koski2016}. In this new point of view, a self-contained universe is composed of a device, the thermal reservoirs, a work source and an information reservoir. This composite system evolves autonomously and any effective feedback control arises from the interplay of the different subsystems.

In the present work, we focus on a particular kind of feedback control, namely, the microcanonical Szil\'ard engine \cite{Marathe2010,Vaikun2011}. In this setup, the demon performs an energy measurement on a system initially equilibrated with a heat bath. The demon then acts according to the outcome of energy measurement keeping the system {\it isolated} from the reservoir. Effectively, one may think of a demon acting on a system initially prepared in a microcanonical ensemble. The examples of microcanonical Szil\'ard engines available in the literature so far deal with the quasistatic regime, which means zero power extracted by the demon. Hence it is highly desirable to construct examples operating in finite time. 

In the quasistatic regime, the energy extraction is a consequence of an effective symmetry breaking \cite{Parrondo2001,Roldan2014,Parrondo2015a}. Nevertheless, we show that it is also possible to extract energy in finite time without splitting the phase space in two or more disconnected regions. Moreover, we show that for certain protocols the work performed in finite time can be equal to the quasistatic work. These are essentially the basic ingredients we use to construct microcanonical Szil\'ard engines producing finite power.

The outline of the paper is as follows: in Sec.~\ref{sec:MSE} we discuss which are the essential features of our construction of microcanonical Szil\'ard engines; in Sec.~\ref{sec:model}, these features are illustrated with an exactly solvable model; an interpretation of our results in terms of phase space is given in Sec.~\ref{sec:Phase_space}; additional examples and final remarks are presented in Sec.~\ref{sec:Nonlinear} and \ref{sec:conclu}, respectively.

\section{\label{sec:MSE}Microcanonical Szil\'ard engines in finite time}

It will be shown in Sec.~\ref{sec:model} and \ref{sec:Nonlinear} that, for a certain class of systems, it it possible to design {\it finite-time} cyclic processes to extract energy from a single heat bath on average. Thus, any intermediate step in the cycle occurs in finite time. Besides, such cyclic processes require {\it feedback}, i.e., they depend crucially on the information gathered by the external observer or demon via an {\it energy} measurement. In particular, it will be shown that these cycles are two-stage processes such that each stage is a {\it linear} variation of a single external control parameter taking time intervals $\tau_{1}$ and $\tau_{2}$. The energy extraction is possible only when these two switching times are carefully chosen based on the outcome of the energy measurement. In Sec.~\ref{sec:model}, we show how to predict these values of $\tau$ for an exactly solvable model using a linear response approach for the excess work (see Eqs.~(\ref{excess_work}) and (\ref{exc_SuperHarm})). In Sec.~\ref{sec:Nonlinear}, we claim that $\tau_{1,2}$ can also be obtained from numerical simulations of non-cyclic process performed on anharmonic oscillators. 

We shall first describe the general features of the cyclic processes we are going to perform. Firstly, the system representing the engine is put in contact with a reservoir at temperature $T$. After its relaxation, the system is found in a Boltzmann-Gibbs distribution $\rho_{eq}(\Gamma,\lambda) = \exp(-\beta \mathcal{H}(\Gamma; \lambda)) / \mathcal{Z}(\beta, \lambda)$, where $\Gamma$ is a point in phase space, $\lambda$ is an external control parameter, $\mathcal{H}$ is the system's Hamiltonian, $\beta = (k_{B}T)^{-1}$, with $k_{B}$ being the Boltzmann constant. The quantity expressed by $\mathcal{Z}(\beta, \lambda)$ is the partition function of the system. 
	
The engine is then decoupled from the heat bath and the demon measures its energy, causing the collapse of the initial Boltzmann-Gibbs distribution into a microcanonical one. After that, the engine will be driven by the demon in a two-stage process while it is kept {\it isolated} from the reservoir. We denote by step 1 the part of the cyclic process in which the system starts with a well-defined energy $E_{1}$ and the external parameter is driven in {\it finite time} from $\lambda_{1}$ to $\lambda_{2}$. This process takes a time interval $\tau_1$ that has been carefully chosen by the demon based on the value of $E_{1}$. To complete the cycle, the external parameter is driven in the opposite direction, i.e., from $\lambda_{2}$ to $\lambda_{1}$. This step 2 takes a time interval $\tau_2$ that is different from $\tau_{1}$ but that also depends crucially on the energy measured by the demon. 
	
\begin{figure}
\includegraphics[width=.4\textwidth]{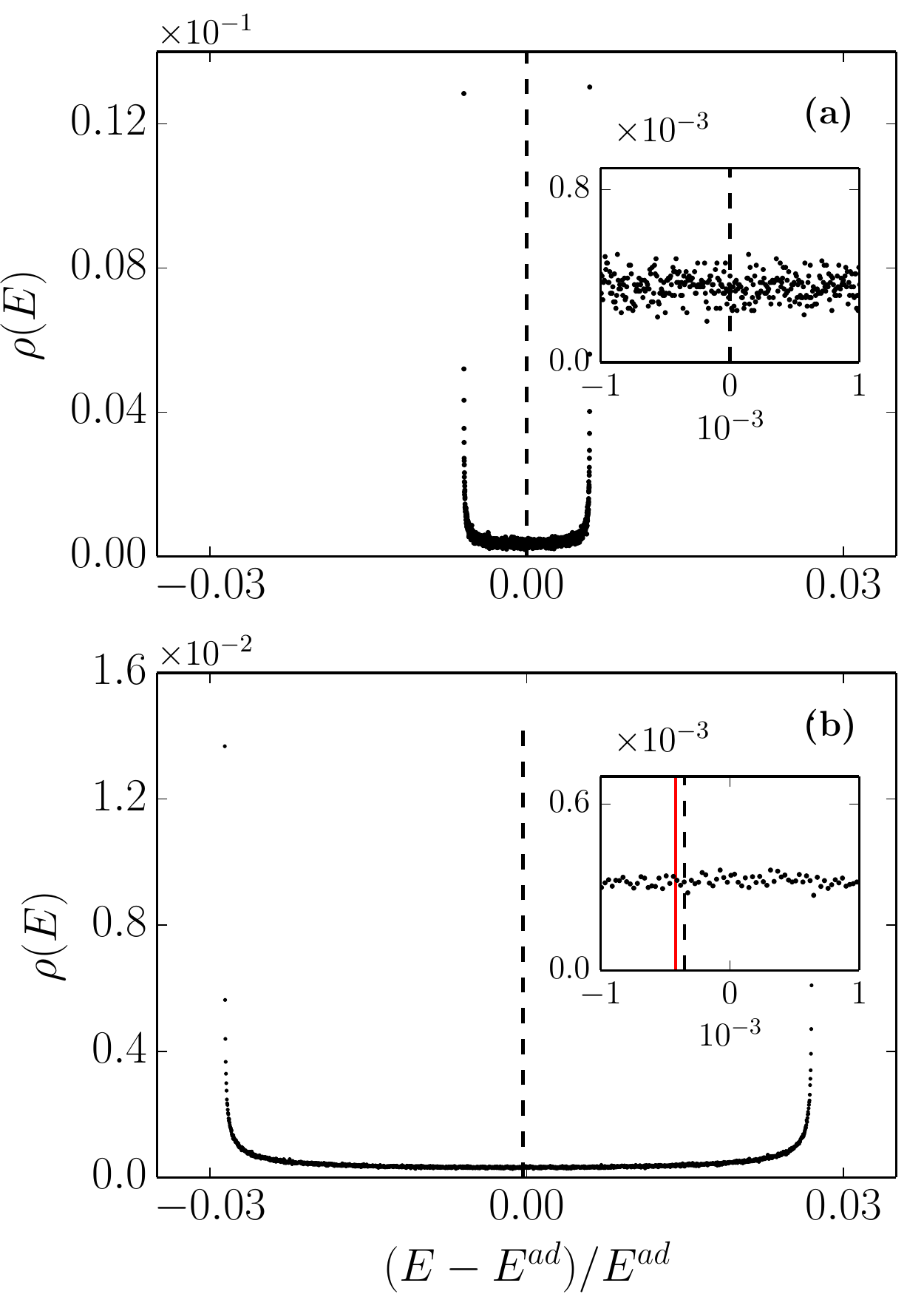}
\caption{(color online) Energy distributions after a finite-time driving of system (\ref{Hamiltonian}) using the linear protocol $\lambda(t) = \lambda_{0}+\delta\lambda(t-t_{0})/\tau$ with $\delta \lambda/\lambda_{0} = 0.1$. Vertical dashed lines represent the average energy for $10^{6}$ initial conditions. The energy was rescaled in terms of the $E^{ad}$, the energy corresponding to the quasistatic evolution of the protocol (see Appendix \ref{AppAI}). (a) Energy distribution after a single linear switching such that $W_{ex} = 0$ (with $\omega\tau \approx 1.8$, see Fig.~\ref{Wex_Super}). (b) Energy distribution after a cyclic switching of $\lambda$ in two linear steps, first, with $W_{ex}=0$ ($\omega\tau_1 \approx 1.8$) and after with $W_{ex} < 0$ ($\omega\tau_2 \approx 2.4$, see Fig.~\ref{Wex_Super}). The values of $\omega\tau_{1,2}$ were calculated using Eq.~(\ref{exc_SuperHarm}). The continuous vertical red line corresponds to the linear response prediction for the average work along the cycle.}
\label{full_cycle}
\end{figure}

In step 1, the demon chooses a particular protocol and finite switching time $\tau_1$ so that the average work performed is the \textit{quasistatic} work $W_{qs}$. In other words, after several realizations of the cycle, the work performed along this stage is on average equal to the value obtained after a quasistatic switching from $\lambda_{1}$ to $\lambda_{2}$. This might sound wrong because we usually think that only quasistatic processes yield work equal to the quasistatic value. However, it has already been discussed elsewhere \cite{Acconcia2015a,Acconcia2015b} that this is not always the case. A numerical evidence of this fact is presented in Fig.~\ref{full_cycle}(a). There, the energy distribution after a finite-time process is shown to give an average value exactly equal to the energy predicted by the adiabatic invariant (see Appendix \ref{AppAI}) for a quasistatic process. We give more details in the next section.
	
In step 2, the demon chooses another particular protocol and finite switching time $\tau_{2}$ to drive $\lambda$ from $\lambda_{2}$ to $\lambda_{1}$. This process is such that the average work $W$ is equal to the negative of the quasistatic work $W_{qs}$ obtained in step 1 plus the excess work $W_{ex}$ (see, for instance, Ref.~\cite{Acconcia2015a} for a discussion about $W_{ex}$). The protocol and switching time $\tau_{2}$ are such that $W_{ex} < 0$ (see Fig.~\ref{Wex_Super} and Sec.~\ref{sec:Nonlinear} for examples). Hence the net work that has been performed in the cycle is negative, meaning that energy was extracted, $W_{cycle}$ = $W_{ex}<0$ (see Fig.~\ref{full_cycle}(b)). It is noteworthy that, in contrast to the corresponding quasistatic engines \cite{Marathe2010,Vaikun2011}, no ergodicity breaking is necessary, i.e., it is not necessary to split the phase space into disconnected regions. It will become clear in what follows how the information gathered by the demon determines the values of $\tau$ for which the above mentioned features happen for each part of the cycle.

In summary, our finite-time microcanonical Szil\'ard engine is based on the existence of {\it finite}-time protocols leading to $W=W_{qs}$ and $W_{ex}<0$ according to the value of the switching time. It remains to be explained how it is possible to find such protocols and why (or more precisely for which class of systems) they yield such features. Based on extensive numerical investigation, we claim that several anharmonic oscillators of one degree of freedom share these properties (see Sec.~\ref{sec:Nonlinear}). In the next section, we take an analytically solvable example to show how to find the protocols we are interested in. It will be shown that all we have to do is to study the behavior of the excess work, $W_{ex}\equiv W - W_{qs}$, as a function of switching time $\tau$ for {\it non}-cyclic processes. Every time $W_{ex}$ is zero or negative for a {\it finite} $\tau$, we have found protocols for step 1 and 2 respectively.

\section{\label{sec:model}Exactly solvable model}

We will consider in this section the following system 
\begin{equation}\label{Hamiltonian}
\mathcal{H}(t) = \epsilon^{-1}\ H_{HO}^{2} = \dfrac{1}{\epsilon} \left[ \dfrac{p^{2}}{2m} + \lambda(t) \dfrac{x^{2}}{2} \right]^{2} \ ,
\end{equation}
where $H_{HO}$ is the harmonic oscillator Hamiltonian, $\epsilon$ is a constant and $\lambda(t)$ is the externally controlled parameter whose time-dependence is expressed in general by $\lambda(t) = \lambda_{0} + \delta\lambda \ g(t)$, where $g(t)$ is such that $g(t_{0}) = 0$ and $g(t_{f}) = 1$, $\tau \equiv t_{f} - t_{0}$. The solutions of  Hamilton's equations are

\begin{subequations}
\begin{align}
x(t) &= x(t_{0}) \cos{(\omega t)}+\frac{p(t_{0})}{m\omega_{0}} \sin{(\omega t)}\ , \\
p(t) &= p(t_{0}) \cos{(\omega t)} - m\omega_{0} x(t_{0}) \sin{(\omega t)}\,, 
\end{align}
\label{coordmom}
\end{subequations}
where $\omega_{0} = \sqrt{\lambda_{0}/m}$ and $\omega = 2\omega_{0}\sqrt{E/\epsilon}$ is the natural frequency of oscillations for an energy $E$. They describe an oscillatory motion with an energy-dependent angular frequency and whose phase space is bounded as, for instance, a pendulum. 

The thermodynamic work $W$ performed when the external parameter $\lambda$ is switched from $\lambda_{0}$ to $\lambda_{f}=\lambda_{0}+\delta\lambda$ reads
\begin{equation}\label{total_work}
W = \int_{t_{0}}^{t_{f}} dt\, \dot{\lambda} \,\overline{\dfrac{\partial \mathcal{H}}{\partial \lambda}} \,,
\end{equation}
where $\overline{A}$ is the out-of-equilibrium average of the quantity $A$.

Restricting ourselves to processes in which $\delta\lambda/\lambda_{0}\ll 1$ (but which are not necessarilly slow), we can apply linear response theory to relate the out-of-equilibrium average $\overline{\partial\mathcal{H}/\partial\lambda}$ to its corresponding relaxation function $\Psi_0(t)$ \citep{Kubo1,Kubo2}. For a system initially in a microcanonical equilibrium distribution whose energy is $E_0=\mathcal{H}(t_0)$, $\Psi_{0}(t)$ is given by \citep{Bonanca2012}
\begin{equation}\label{relax_func}
\Psi_0(t) = \dfrac{1}{Z(\lambda_{0},E_0)} \dfrac{\partial }{\partial E_0} \left[ Z(\lambda_{0},E_0)(C(t) - \mathcal{C})\right] \ ,
\end{equation}
where $Z(\lambda_{0},E_0) = \int dx\,dp\,\delta(E_{0}-\mathcal{H}(x,p;\lambda_{0}))$, $C(t) = \langle \partial \mathcal{H}(0)/\partial\lambda \ \partial \mathcal{H} (t)/\partial\lambda \rangle_{0}$ and $\mathcal{C}$ is the asymptotic value, for large $t$, of the correlation function $C(t)$ (for simply oscillatory functions, this has to be properly defined. See, for instance, Sec. 4.2.2 of \cite{Kubo2}). We denote by $\langle A \rangle_{0}$ the microcanonical average of the observable $A$.

After some algebra linking Eq.~(\ref{total_work}) to the relaxation function (\ref{relax_func}) (see Appendix \ref{Appendix_A}), we obtain an expression for $W$ given by the sum of two contributions. The first one can be identified with the quasistatic work along a equivalent quasistatic process and reads 
\begin{equation}\label{qs_work}
W_{qs} = \delta\lambda \left\langle\dfrac{\partial \mathcal{H}}{\partial\lambda} \right\rangle_{0} - \dfrac{(\delta \lambda)^{2}}{2} \tilde{\Psi}_{0}(0) \ ,
\end{equation}  
where $\tilde{\Psi}_{0} \equiv \Psi_{0}(0) - \chi_{0}^{\infty}$ with $\chi_{0}^{\infty} = \langle \partial^2 \mathcal{H}/\partial \lambda^2 \rangle_0$.

It is worth notice that both terms \textit{do not} depend on the protocol $g(t)$. Indeed, it can be verified that they are the first terms of the series expansion of the quasistatic work for $\delta\lambda/\lambda_{0} \ll 1$. 

The second contribution for the total work is a term that vanishes in the quasistatic limit and clearly depends on $g(t)$. Therefore, it is called excess work and reads \cite{Acconcia2015a},
\begin{equation}\label{excess_work}
W_{ex} = \frac{(\delta\lambda)^{2}}{2}\int_{t_{0}}^{t_{f}}dt \int_{t_{0}}^{t_{f}}dt'\,\Psi_{0}(t-t') \dot{g}(t) \dot{g}(t')\,.
\end{equation} 
In other words, $W_{ex} \equiv W-W_{qs}$ is the extra amount of energy the external agent has to pump into the system during a finite-time process. 

The previous analysis about the linear response expression for $W$ is very general and can now be applied to system (\ref{Hamiltonian}). The correlation function $C(t)$ can be calculated analytically from the previous expressions for $x(t)$ and $p(t)$, Eq.~(\ref{coordmom}), yielding the following relaxation function
\begin{equation}\label{relax_func_harm}
\Psi_{0}(t) = \dfrac{E}{4 \lambda_{0}^{2}}\left[  3 \cos(2\omega t) - 2 \omega t \sin(2\omega t)\right].
\end{equation} 

\begin{figure}
\centering
\includegraphics[width=.43\textwidth]{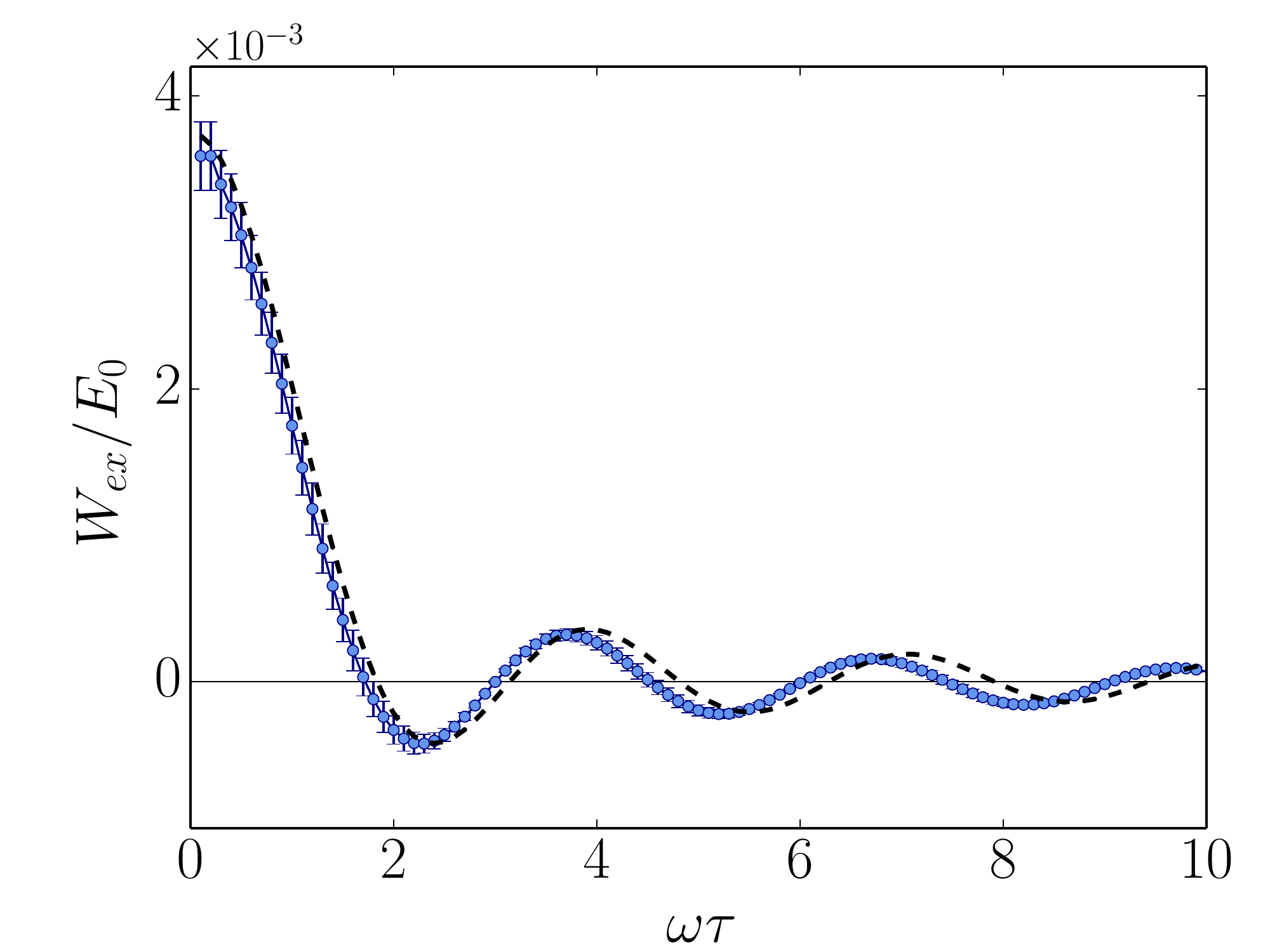} 
\caption{(color online) Comparison between numerical results (blue circles) and linear response prediction (Eq.~(\ref{exc_SuperHarm})) (dashed line) of $W_{ex}$ as a function of the switching time $\tau$ for system (\ref{Hamiltonian}). It was chosen the linear protocol $g(t)=(t-t_0)/\tau$ and $\delta\lambda/\lambda_0 = 0.1$. We used $10^{6}$ initial conditions to calculate numerically the excess work for each $\tau$.}
\label{Wex_Super}
\end{figure}

Finally, we can calculate $W_{ex}$ analytically for a given protocol $g(t)$ using Eq.~(\ref{excess_work}). In particular, for the linear protocol $g(t) = (t-t_0)/\tau$ the result reads
\begin{eqnarray}\label{exc_SuperHarm}
\lefteqn{W_{ex}(\tau) =}\nonumber \\
&& \dfrac{E_0}{8} \left( \dfrac{\delta\lambda}{\lambda_{0}}\right)^{2}\dfrac{\sin(\omega\tau)[(2\omega\tau)\cos(\omega\tau) + \sin(\omega\tau)]}{(\omega\tau)^{2}}.
\end{eqnarray}
The previous expression has several important features. As mentioned before, it goes to zero when we take the limit $\omega\tau \rightarrow \infty$. For large values of $\omega\tau$, we find that $W_{ex} \propto 1/\tau$, which can be related to the Sekimoto-Sasa relation presented in Refs.\cite{Sekimoto1997,Roldan2017}. Equation (\ref{exc_SuperHarm}) also reveals that $W_{ex}$ can be null for \textit{finite} values of $\tau$. This happens either (i) when $\sin{(\omega\tau)}$ is zero, i.e., $\omega\tau = 2\pi k$, with $k$ an integer; or (ii) when $\tan(\omega\tau)/2 = - \omega\tau$. It also predicts {\it negative} values of $W_{ex}$ for specific ranges of $\tau$. Figure~\ref{Wex_Super} shows the comparison between Eq.~(\ref{exc_SuperHarm}) and numerical simulations using the linear protocol. Numerics were implemented using symplectic integrators \cite{Channel1990}.	
	
\begin{figure}
\includegraphics[scale=0.4]{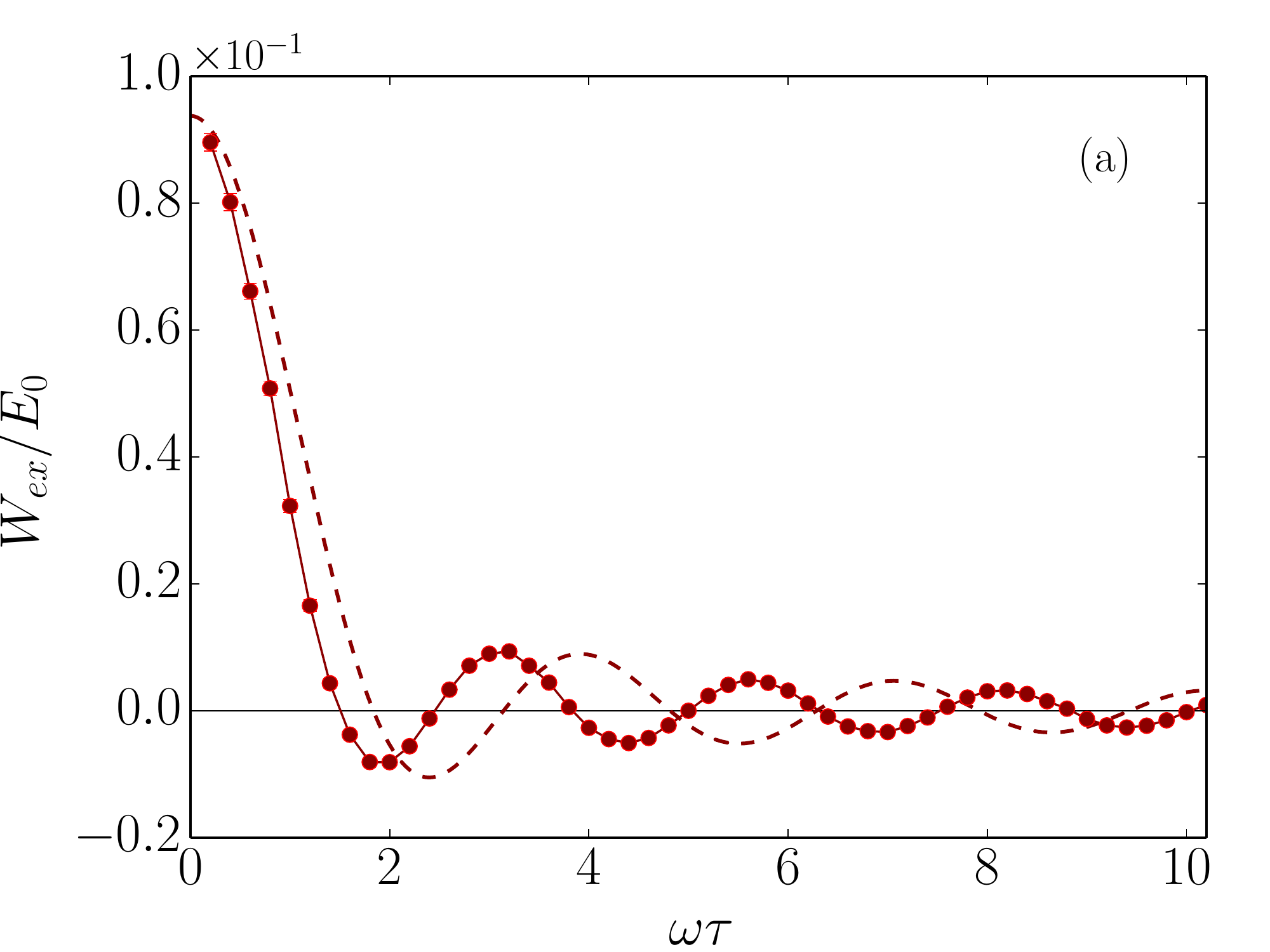}
\includegraphics[scale=0.4]{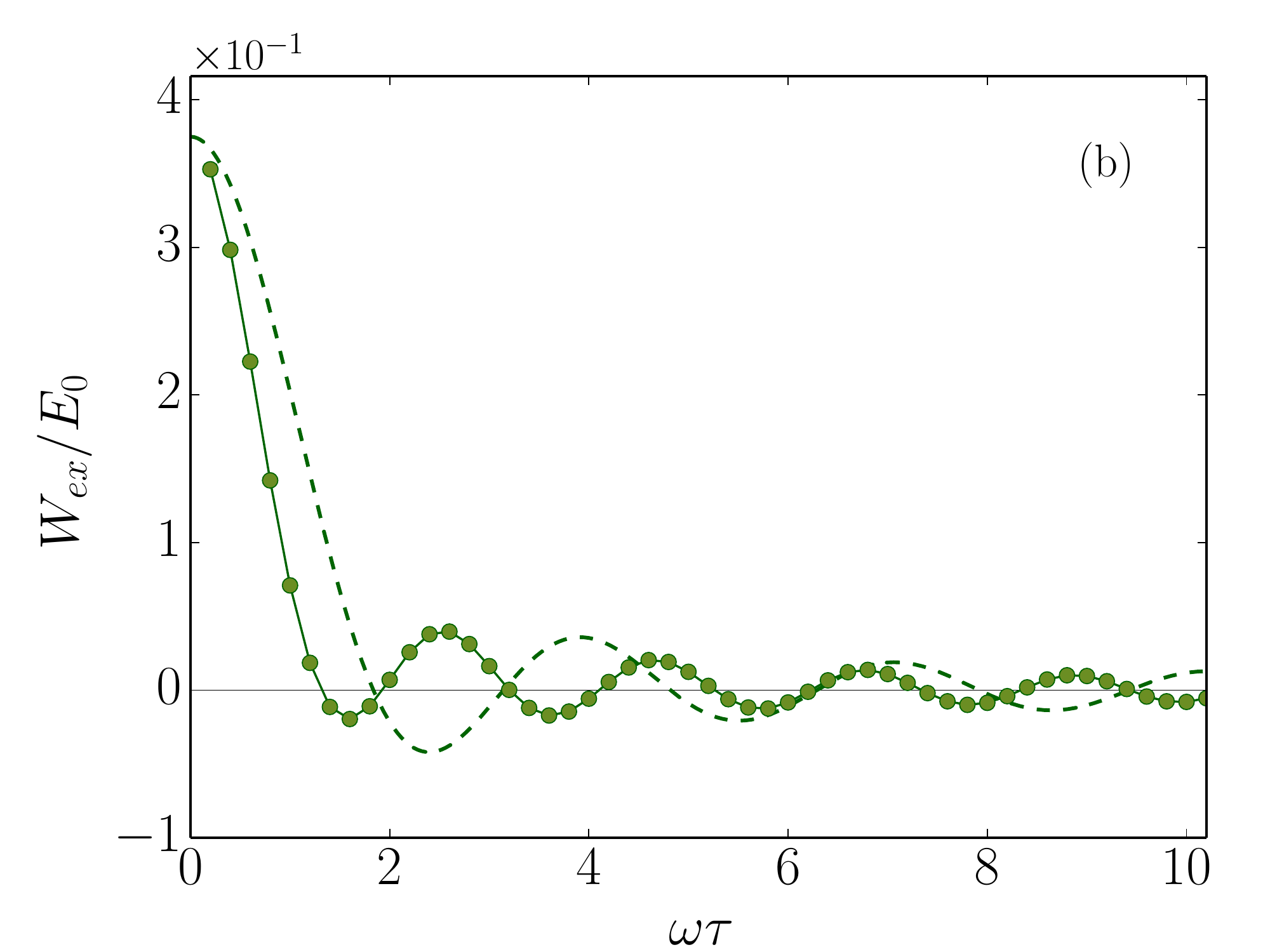}
\caption{(color online) Comparison between numerical calculations (circles) and linear response predictions (Eq.~(\ref{exc_SuperHarm})) (dashed line) of $W_{ex}$ as a function of the switching time $\tau$ for system (\ref{Hamiltonian}). The driving was performed using the linear protocol $g(t) = (t-t_{0})/\tau$ with (a) $\delta\lambda / \lambda_{0} = 0.5$ and (b) $\delta\lambda / \lambda_{0} = 1.0$. We used $10^{6}$ initial conditions to calculate numerically the excess work for each $\tau$.}
\label{high_ampli}
\end{figure}

The agreement between theoretical prediction for $W_{ex}$ and the numerical calculation is very good. Although the theoretical prediction is restricted to weak driving, i.e., $\delta\lambda/\lambda_0 \ll 1$, it includes arbitrarily fast processes. Besides, Eq.~(\ref{exc_SuperHarm}) predicts the special switching times for which the excess work is either zero or negative. Additionally, we have investigated numerically the behavior of the excess work upto $\delta\lambda/\lambda_{0} = 1.0$. Figure~\ref{high_ampli} shows that the linear response expression clearly deviates from the numerical calculations as $\delta\lambda/\lambda_{0}$ increases. However, it also shows that negative values of excess work are not restricted to our linear response regime. This suggests that our construction of a microcanonical Szil\'ard engine might be valid beyond the weak driving regime. We must be careful about this because, as will be explained next, our proposal depends crucially on how small energy fluctuations are after the first step of the cycle and we have not studied them in detail. 
	
In summary, it is possible to implement a microcanonical Szil\'ard engine driving system (\ref{Hamiltonian}) with linear protocols as follows. In step 1, the control parameter is driven from $\lambda_{0}+\delta\lambda$ to $\lambda_{0}$ in a time interval $\tau_{1}$ yielding the first zero of $W_{ex}$. Since the energy distribution after this step is a very narrow one (see Fig.~\ref{full_cycle}), we assume that the initial energy distribution for step 2 is again a microcanonical one. Hence the linear response results for $W_{ex}$ can be used to predict what is going to happen in both steps of the cycle. This is the reason why we first perform a process with $W_{ex}=0$. Although the inset in Fig.~\ref{full_cycle}b shows that this is indeed a good approximation, this might be not the case beyond the linear response regime. In step 2, another linear protocol drives $\lambda$ back to $\lambda_{0}$ taking a time interval $\tau_{2}$ such that $W_{ex} < 0$. Thus, the net work performed in the cycle is $W_{ex}(\tau_{2})$. In Fig.~\ref{full_cycle}(b), we show the numerical result for the average energy after this cycle. Since $\omega$ is a function of the energy, the duration of each protocol is chosen based on the information gathered by demon about the initial energy. 

In contrast to the quasistatic regime, the finite-time driving of the system leads to different values of work after each single realization of the protocol. Hence we want to stress that our results are valid on average. Indeed, every time the demon starts a new cycle, the outcome of the energy measurement differs from the previous one. To obtain the curve in Fig.~(\ref{Wex_Super}), the demon has then to sort his ensemble of trajectories by their initial energy values and take an average using only trajectories with essentially the same initial energy. Instead, if all trajectories are considered, the work performed in a cycle is on average equal to

\begin{eqnarray}\label{canonical_average}
\langle W_{cycle}\rangle &=& \dfrac{1}{\mathcal{Z}(\beta)} \int_{0}^{\infty} dE \ e^{-\beta  E} Z(E) \ W_{ex}(E,\tau_2(E)) \nonumber \\
&=& - \dfrac{1}{16\beta} \left( \dfrac{\delta\lambda}{\lambda_0}\right)^{2} \vert f(\omega\tau_2) \vert \ ,
\end{eqnarray}
where $W_{ex}(\tau_2)$ is the value of the first minimum of Eq.~(\ref{exc_SuperHarm}) , $f(\omega\tau) = \sin(\omega\tau)[2\omega\tau \cos(\omega\tau) + \sin(\omega\tau)] / (\omega\tau)^{2}$ and $Z(E)$ is the density of states. Since $\omega\tau_{2} \approx 2.4$ does not depend on the energy, $f(\omega\tau_2) \approx -0.34$ can be taken outside the integral. It is worth mentioning that the demon can always find a suitable protocol yielding $W_{ex} < 0$ no matter the value of the energy measured.

\section{\label{sec:Phase_space}Phase space dynamics}

After presenting the linear response description of finite-time microcanonical Szil\'ard engines, we shall discuss the physical mechanism behind the energy extraction in our setup. The protocols we have discussed previously never split the phase space of the system in two or more disconnected parts \cite{Roldan2014,Parrondo2001,Parrondo2015a}. Nevertheless, Figure~\ref{full_cycle} shows that, after a finite-time process, the energy distribution is essentially zero except for two particular values of energy, i.e., the system basically assumes either one or the other value of energy. 

\begin{figure}
\centering
\includegraphics[width=.45\textwidth]{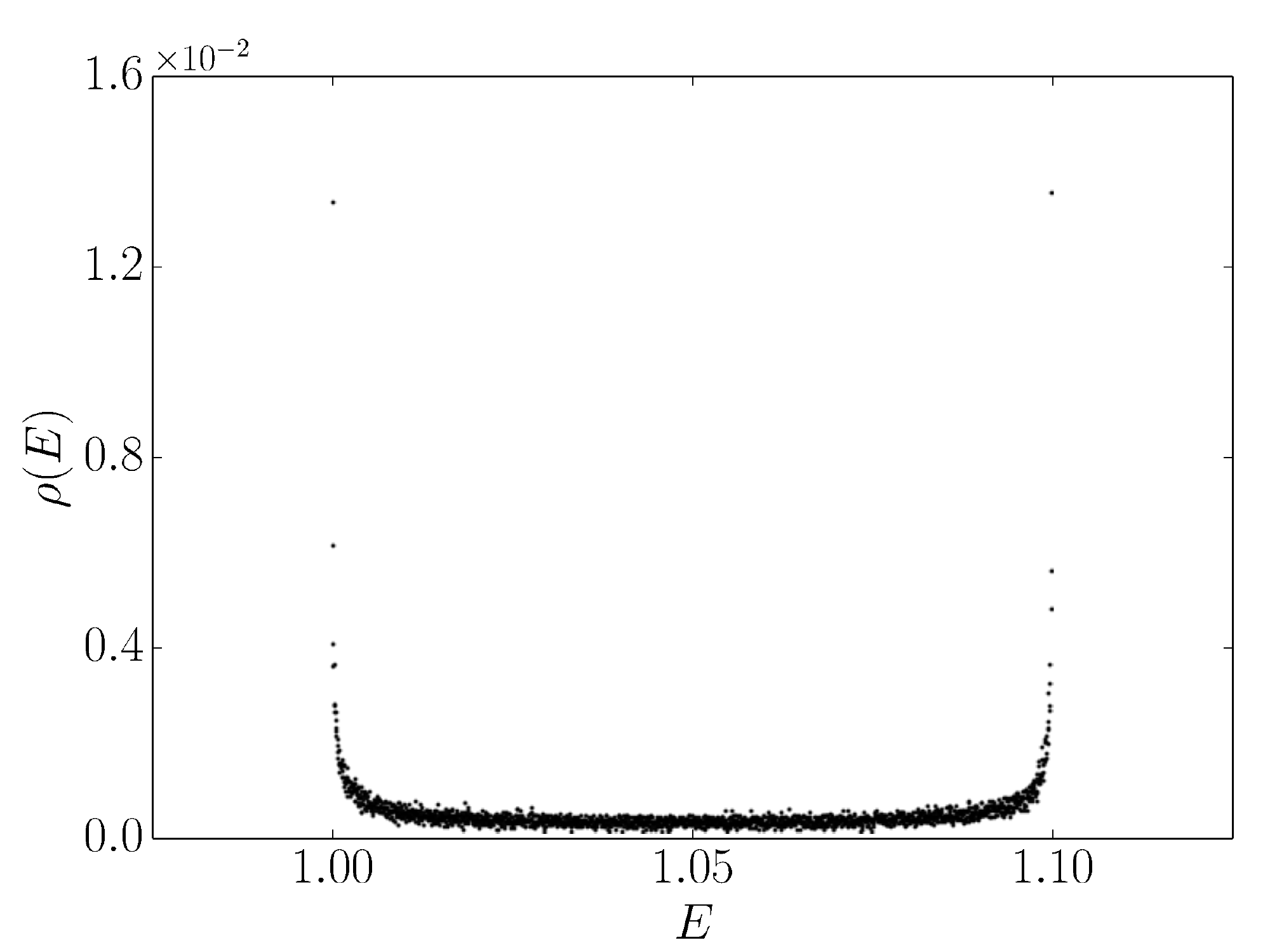}
\caption{(color online) Energy distribution of the harmonic oscillator (\ref{harmosc}) after a finite-time linear switching of $\lambda$. We set $\delta\lambda / \lambda_{0} = 0.1$ and $\tau = 0.1\omega_0^{-1}$. We used $5\times 10^{4}$ initial conditions.}
\label{Energy_Harm_osc}
\end{figure}

\begin{figure}
\centering
\includegraphics[width=.46\textwidth]{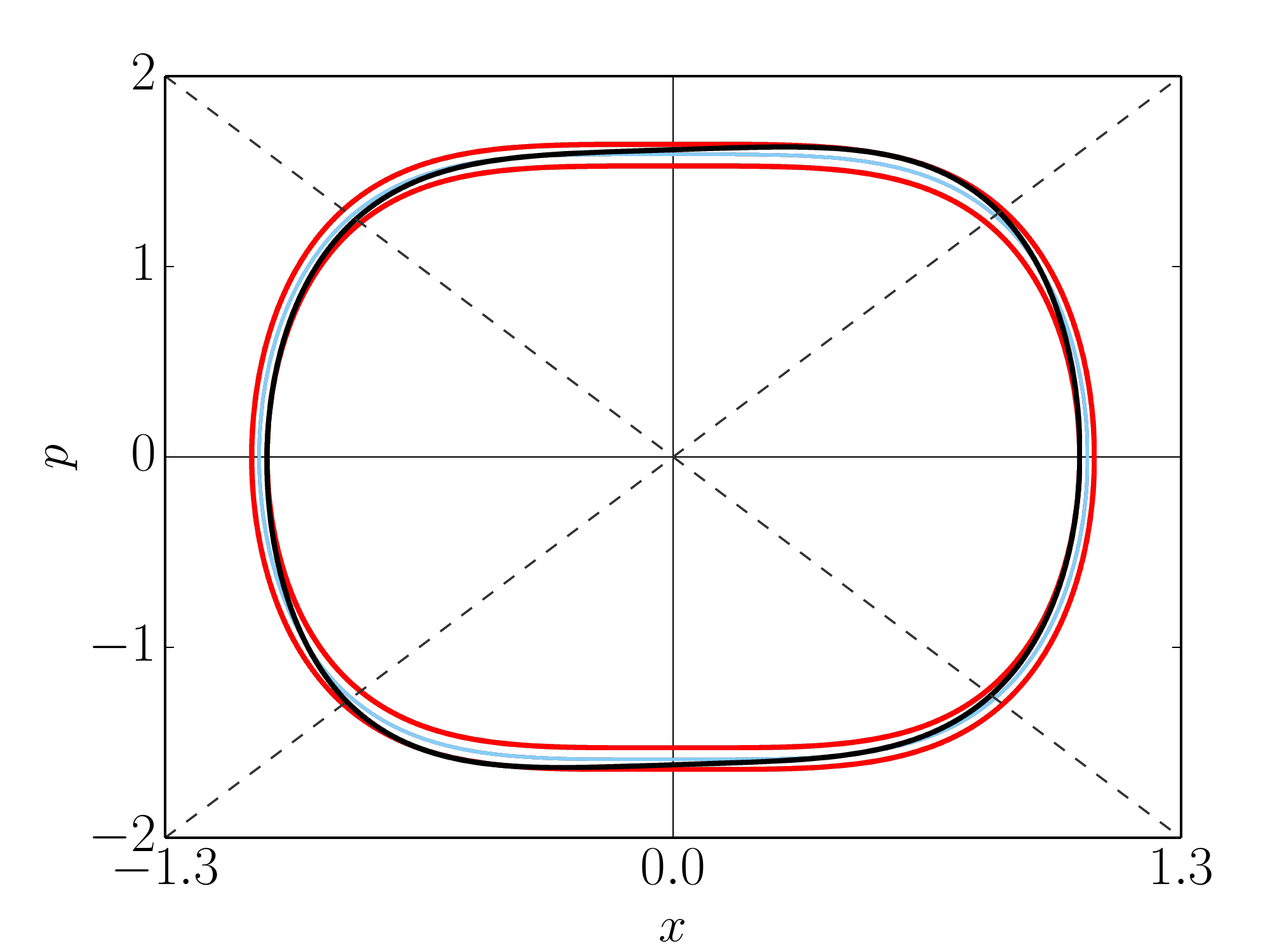} 
\caption{(color online) Phase space distribution for the quartic oscillator, Eq.~(\ref{quartic}). The black curve represents the deformed energy shell after the finite-time protocol yielding $W_{ex} < 0$ (see Fig.~\ref{distrib_quartic}). The outer and inner red (dark gray) curves are the energy shells corresponding, respectively, to the maximum and minimum values of the energy distribution (see Fig.~\ref{distrib_quartic}) after the following finite-time protocol: $g(t) = (t-t_{0})/\tau$, $\delta\lambda/\lambda_{0} = 1.0$, $\lambda_{0}=1.0$ and $W_{ex}(\tau)<0$. The blue (light gray) curve corresponds to the energy shell obtained after the corresponding quasistatic driving.}
\label{phase_space}
\end{figure}

It is necessary to demand a bit more from the energy distribution in order to extract energy. The balance between the two peaks of the distribution has to be such that the average energy is \textit{smaller} than the energy obtained after an equivalent quasistatic process, otherwise $W_{ex}$ never becomes negative. This seems to be impossible for a harmonic oscillator as shown below. Its energy distribution after a finite-time driving also presents two pronounced peaks (see Fig.~\ref{Energy_Harm_osc}) but $W_{ex}$ is always non-negative (see Eq.~(\ref{nonneg_ex}) and Fig.~\ref{curve_Harm_osc}). Our claim about why the anharmonic oscillators yield $W_{ex} < 0$ is based on the phase-space plots in Fig.~\ref{phase_space} for the quartic oscillator of Eq.~(\ref{quartic}).

The initial energy shell is deformed along the momentum and coordinate directions when the system is driven quasistatically. Nevertheless, our numerical simulations indicate that the stretching and contraction directions rotate when an anharmonic oscillator is driven in finite time. This happens in such way that the portions of the deformed curved with energy below the energy after the quasistatic process are slightly larger than those with energy higher than that.
	
\subsection{\label{Harm_osc}Harmonic Oscillator}

We briefly discuss now why the harmonic oscillator does not reproduce the behavior we observe in Fig.~\ref{Wex_Super}. For the Hamiltonian,
\begin{equation}
H_{HO} = \dfrac{p^{2}}{2m} + \lambda(t) \dfrac{x^{2}}{2} \ ,
\label{harmosc}
\end{equation}
the relaxation function can be obtain as described previously and it reads
\begin{equation}
\Psi_{0}(t) = \dfrac{E}{4\lambda_{0}^{2}}  \cos(2\omega_{0} t)\,.
\label{relaxharm}
\end{equation}

Equations~(\ref{excess_work}) and (\ref{relaxharm}) then yield
\begin{eqnarray}
W_{ex} &\propto& \int_0^1 ds \int_0^1   du \ \cos(2\omega_0\tau(s-u)) \dot{g}(s) \ \dot{g}(u) \nonumber \\
&=& \left( \int_0^1 ds\  \cos(2\omega_0\tau s) \ \dot{g}(s) \right)^2  + \nonumber \\
&+&  \left( \int_0^1 du \ \sin(2\omega_0\tau u) \ \dot{g}(u) \right)^2 \,.
\label{nonneg_ex}
\end{eqnarray}

\begin{figure}
\centering
\includegraphics[width=.45\textwidth]{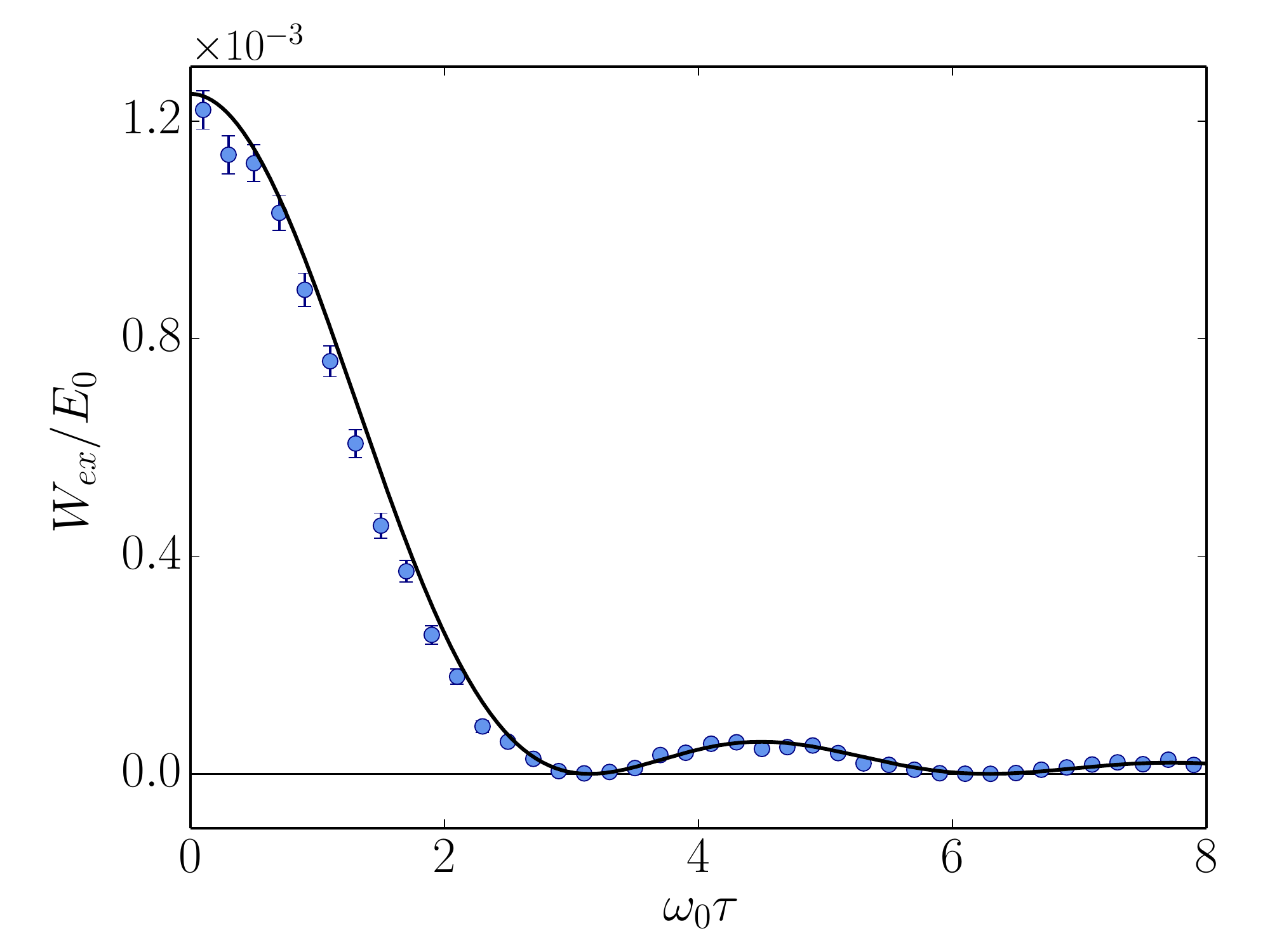}
\caption{(color online) Excess work as a function of the switching time for the harmonic oscillator, Eq.~(\ref{harmosc}). We compare numerical calculations (blue circles) and linear response prediction (solid line) when $g(t) = (t-t_{0})/\tau$. We used $10^{6}$ initial conditions for each switching time $\tau$ and fixed $\delta\lambda / \lambda_{0} = 0.1$. }
\label{curve_Harm_osc}
\end{figure}

The excess work is therefore always positive for the harmonic oscillator. In other words, we could use this system to obtain zero excess work in finite time but it would be impossible to model a microcanonical Szil\'ard engine with it along the same lines presented previously. Figure~\ref{curve_Harm_osc} shows the comparison between numerical calculations and the linear response expression for the linear protocol.
	
Figure~\ref{Energy_Harm_osc} shows the energy distribution of the harmonic oscillator after a linear switching of $\lambda$. Although it presents the same structure we find in anharmonic oscillators, it does not favor the lower value of energy and, consequently, it does not give rise to negative values of $W_{ex}$.

\begin{figure}
\centering
\includegraphics[width=.47\textwidth]{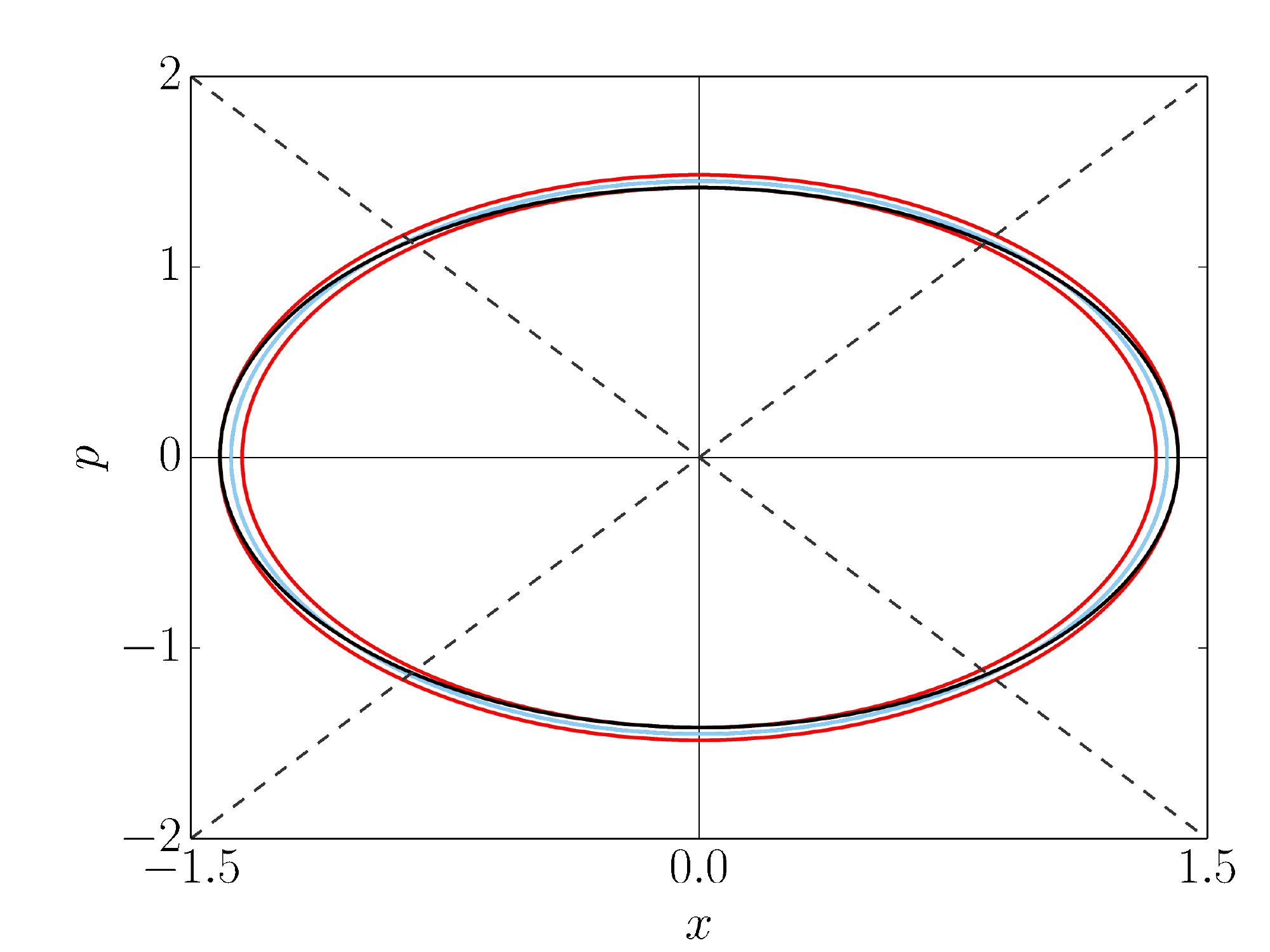}
\caption{(color online) Deformation of the initial energy shell of system (\ref{harmosc}) after a finite-time driving of $\lambda$. The outer and inner red (dark gray) curves are energy shells corresponding, respectively, to the maximum and minimum values of the energy distribution in Fig.~\ref{Energy_Harm_osc}. The black curve represents the finite-time deformation of the initial energy shell after the linear protocol $g(t) = (t-t_{0})/\tau$ with $\delta\lambda/\lambda_{0} = 0.1$ and $\omega_{0}\tau = 0.1$. The blue (light gray) curve corresponds to the energy shell obtained after a quasistatic driving. We used $5\times 10^{4}$ initial conditions for each curve.}
\label{Phase_Harm_osc}
\end{figure}

Figure~\ref{Phase_Harm_osc} shows the phase-space representation of the states leading to the energy distribution in Fig.~\ref{Energy_Harm_osc}. Since the switching time was chosen such that $W_{ex} > 0$, the initial set of phase-space points evolves to a curve that, although very close, is not a energy shell. The initial energy shell deforms along the momentum and coordinate axis even in finite time. This contrasts with what happens in anharmonic oscillators where the stretching and contraction directions rotate and coincide with the momentum and coordinate axis only in the quasistatic regime.

\section{\label{sec:Nonlinear}Other examples}

The behavior of $W_{ex}$ for the system (\ref{Hamiltonian}) is typically found in other anharmonic oscillators. In what follows, we show numerical calculations of $W_{ex}$ for a few examples: the quartic oscillator, 
\begin{equation}
\mathcal{H}[\lambda(t)] = \dfrac{p^{2}}{2m} + \lambda(t) \dfrac{x^{4}}{2} \,,
\label{quartic}
\end{equation}
the pendulum,
\begin{equation}
\mathcal{H}[\lambda(t)] = \dfrac{p^{2}}{2m} + 2\lambda(t) \sin^{2}\left(\dfrac{x}{2}\right) \,,
\label{pendulum}
\end{equation}
and the logarithmic oscillator,
\begin{equation}
\mathcal{H}[\lambda(t)] = \dfrac{p^{2}}{2m} + \lambda(t) \log\left(\dfrac{x^2+b^2}{b^2}\right) \ ,
\label{logarithm}
\end{equation}
where $b$ is a fixed parameter.

\begin{figure}
\centering
\includegraphics[width=.45\textwidth]{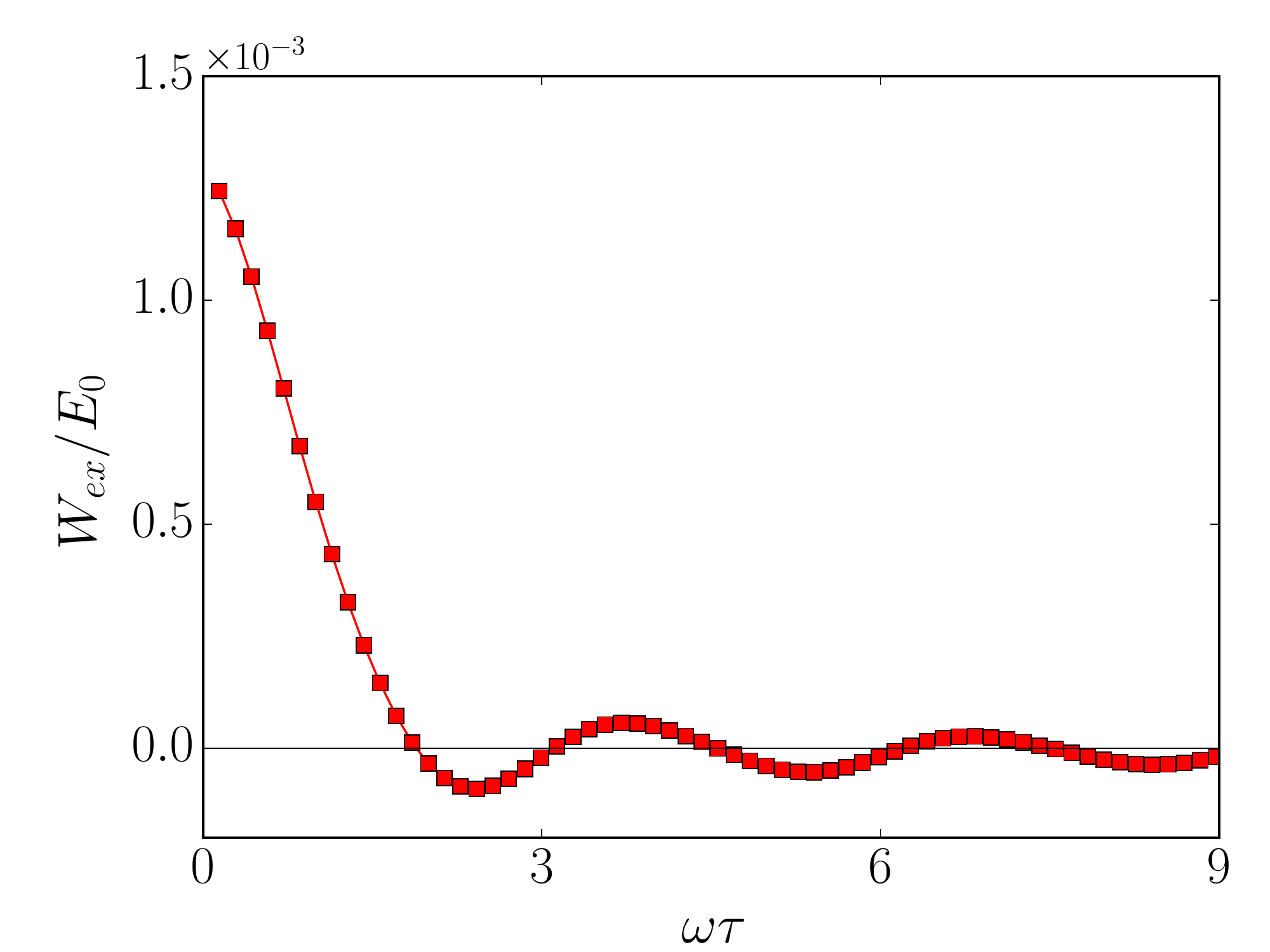}
\caption{(color online) Excess work as a function of the switching time for the system (\ref{quartic}) using the linear protocol $g(t) = (t-t_{0})/\tau$, $\delta\lambda / \lambda_{0} = 0.1$ and $10^{6}$ initial conditions for each value of $\tau$.}
\label{quartic_potential}
\end{figure}

\begin{figure}
\centering
\includegraphics[width=.45\textwidth]{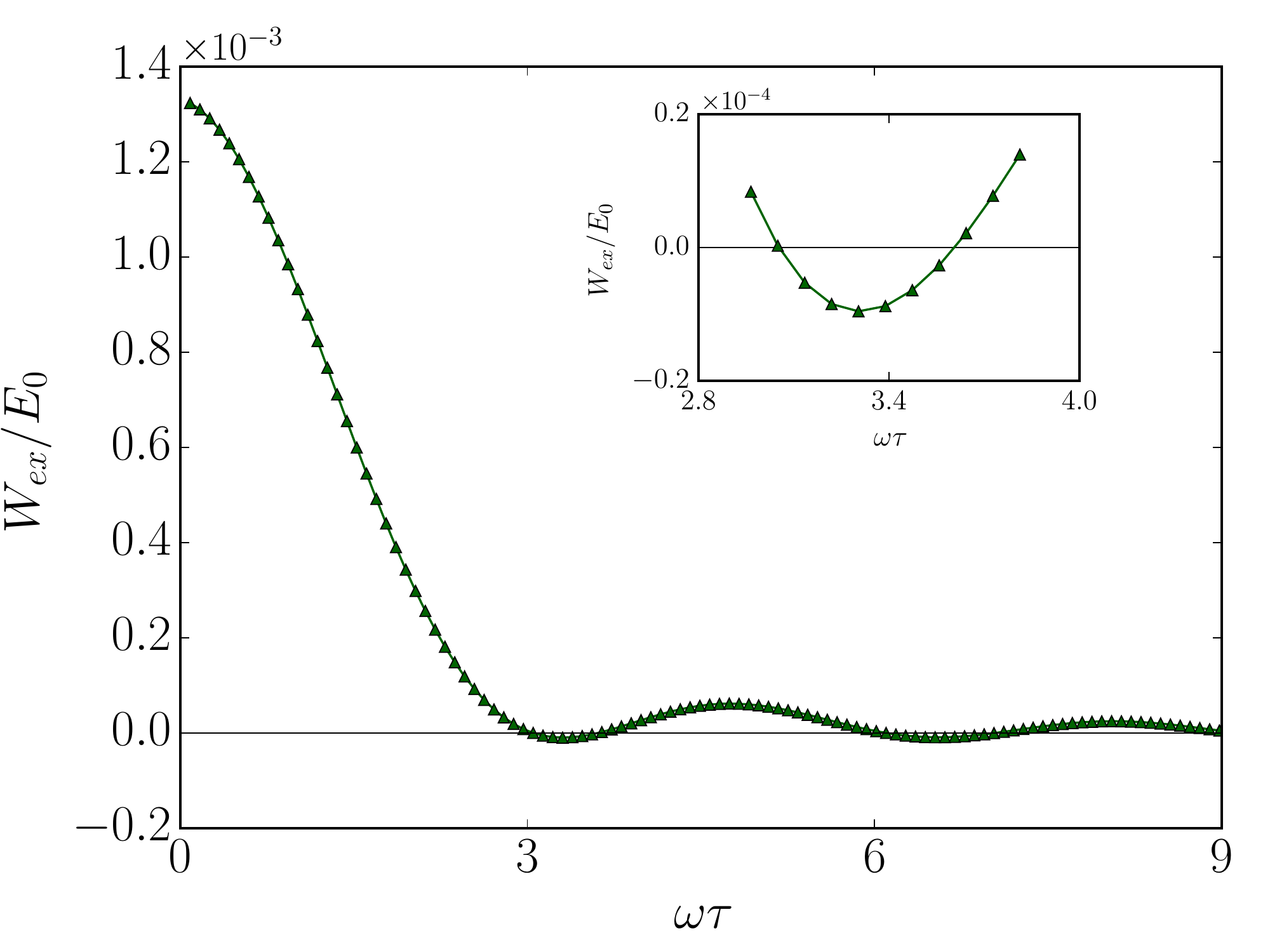}
\caption{(color online) Excess work as a function of the switching time for the system (\ref{pendulum}) using the linear protocol $g(t) = (t-t_{0})/\tau$, $\delta\lambda / \lambda_{0} = 0.1$ and $10^{6}$ initial conditions for each value of $\tau$.}
\label{pendulum_potential}
\end{figure}

\begin{figure}
\centering
\includegraphics[width=.45\textwidth]{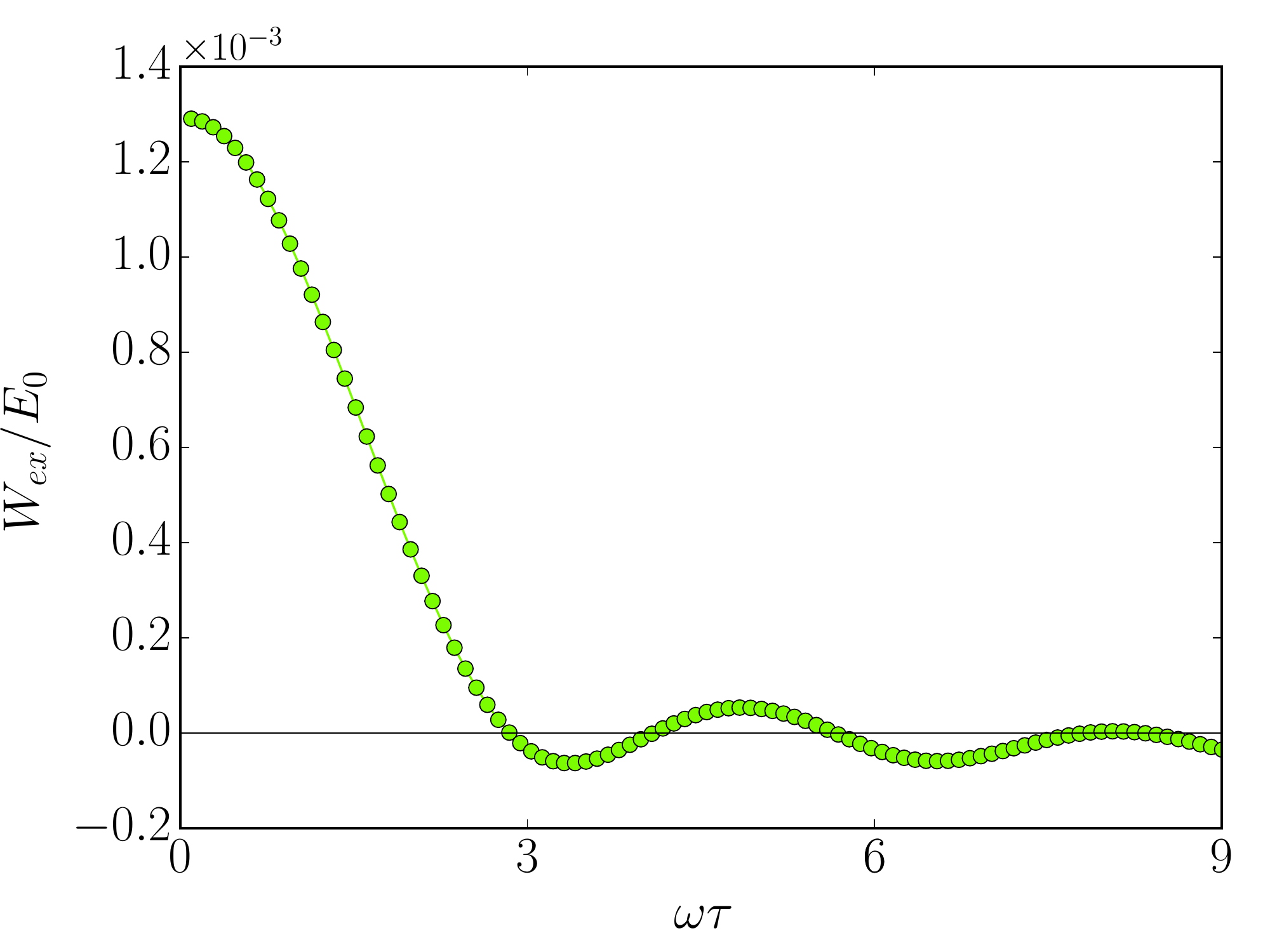}
\caption{(color online) Excess work as a function of the switching time for the system (\ref{logarithm}) using the linear protocol $g(t) = (t-t_{0})/\tau$, $\delta\lambda/\lambda_{0} = 0.1$ and $10^{6}$ initial conditions for each value of $\tau$.}
\label{log_potential}
\end{figure}

\begin{figure}
\centering
\includegraphics[width=.45\textwidth]{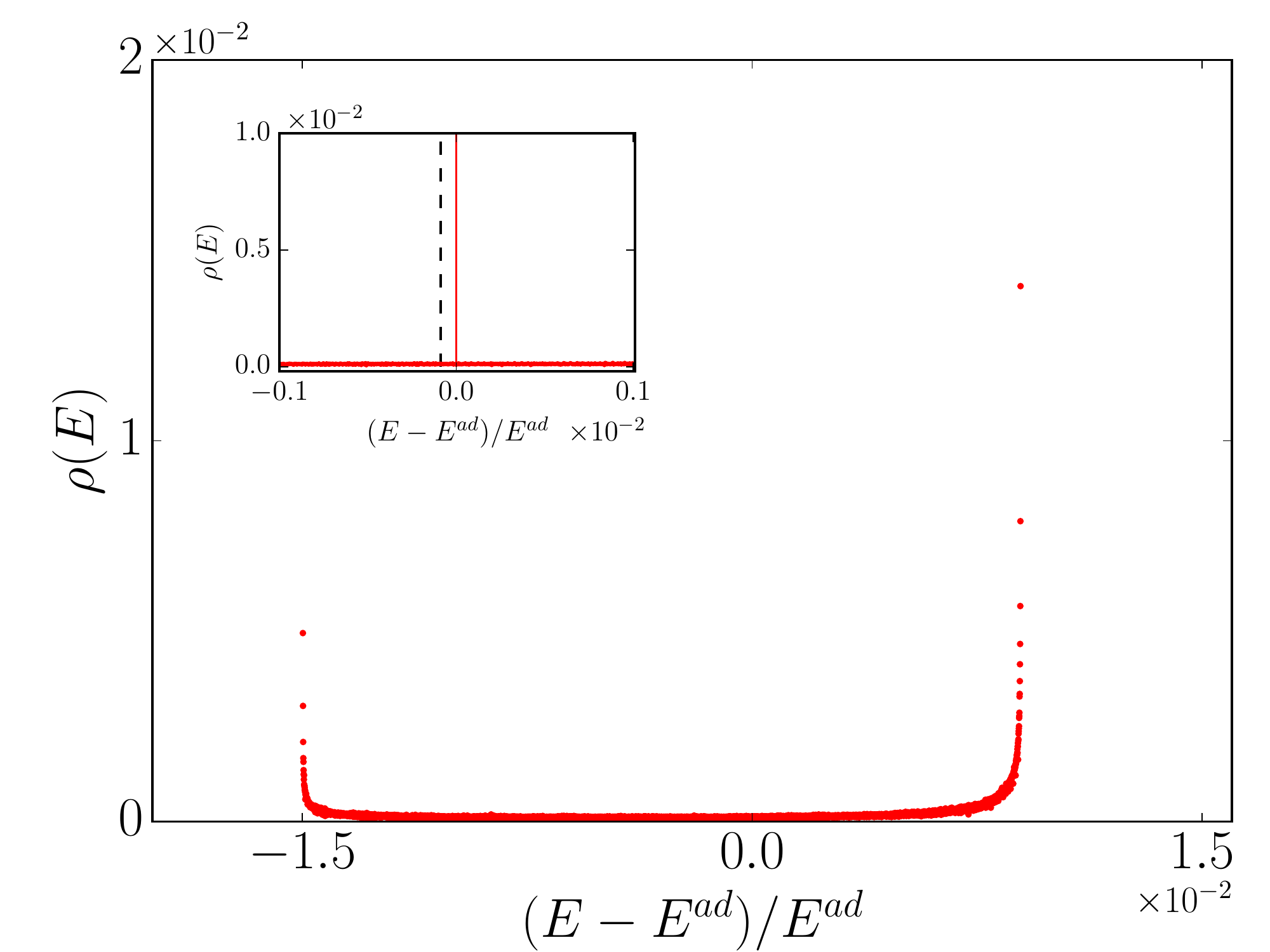}
\caption{(color online) Energy distribution (see Appendix \ref{AppAI} for a definition of $E^{ad}$) after a finite-time driving of system
(\ref{quartic}) using a linear protocol $g(t) = (t-t_{0})/\tau$ such that $\delta\lambda / \lambda_{0} = 1.0$ and $W_{ex}(\tau)< 0$. We used $10^{6}$ initial conditions. The values of energy were rescaled by the corresponding value $E^{ad}$ after a quasistatic driving (see Appendix \ref{AppAI}).}
\label{distrib_quartic}
\end{figure}

The corresponding excess work as a function of the switching time is presented in Figs.~\ref{quartic_potential} to \ref{log_potential} for a linear switching of $\lambda$. We would like to highlight that the system (\ref{logarithm}) was used as an ideal Hamiltonian thermostat in \cite{Campisi2012}. We also present the energy distribution of the system (\ref{quartic}) in Fig.~\ref{distrib_quartic} to stress the average negative excess work after the finite-time driving for the particular switching time $\omega\tau \approx 2.6$.
	
\section{\label{sec:conclu}Concluding Remarks}

In conclusion, we have shown how to construct finite-time microcanonical Szil\'ard engines for systems whose excess work presents finite-time zeros and negative values in non-cyclic processes. The exact values of switching times allowing for such values of excess work have to be carefully chosen based on the information gathered by the demon. In contrast to previous examples available in the literature, these engines produce finite power on average. The cyclic process was obtained using a linear response approach which can be further explored to furnish the optimal protocols yielding the maximum energy extraction. Despite of the unexpected absence of ergodicity breaking, we have shown that it is still possible to extract energy due to special features of the energy distribution. This question deserves further investigation especially if one wants to extend the present results to systems with a larger number of degrees of freedom. 

The energy extraction in Szil\'ard engines is usually attributed to the sudden reduction of phase space volume due to a symmetry breaking \cite{Roldan2014,Parrondo2001,Parrondo2015a}. In the present case, although there is no such mechanism, we may have a reduction of phase space volume on average. Since the volume $\Omega$ enclosed by an energy shell is a monotonic increasing function of the energy $E$, there is a one-to-one mapping between $\Omega$ and $E$. Thus, from a energy distribution as Fig.~\ref{full_cycle}, it is possible to obtained the corresponding distribution of $\Omega$, whose average value is going to be smaller than the initial one every time $W_{ex} < 0$. In this sense, we might connect the energy extraction to a reduction of phase space volume on average. 

As a final remark, we should mention that the quantification of the information gathered my the demon in our examples does not seem as straightforward as in Ref.~\cite{Vaikun2011}. This analysis will be left for a future work. Nevertheless, it is clear that we have shown examples of a feedback process in which the outcome of an energy measurement allows for energy extraction.

\begin{acknowledgments}
The authors acknowledge J. P. Pekola, P. Talkner, M. Campisi, S. Deffner and W.H. Zurek for enriching discussions. We especially thank \'{E}. Rold\'{a}n for the careful analysis of the manuscript and inspiring discussions. T.A. acknowledges support from `Gleb Wataghin' Physics Institute (Brazil) and CAPES (Brazil), Grant No. 1504869. M.B. acknowledges financial support from Unicamp/FAEPEX (Brazil), Grant No. 0031/15 and FAPESP (Brazil), Grant No. 2016/01660-2.
\end{acknowledgments}

\appendix
%%%%%%%%%%%%%%%%%%%%%%%%%%%%%%%%%%%%%%%%%%%%%%%%%%%%%%%%%%%%
\section{Adiabatic invariant}\label{AppAI}

We define here what we mean by $E^{ad}$, the energy we use to rescale the energy distributions in Figs.~\ref{full_cycle} and \ref{distrib_quartic} and to calculate the work $W_{qs}$ along a quasistatic process. Firstly, it is necessary to define the ``volume" $\Omega(E,\lambda)$ enclosed by the surface of constant energy $H(\mathbf{q},\mathbf{p};\lambda) = E$,
\begin{equation}
\Omega(E,\lambda) = \int  d\mathbf{q}\,d\mathbf{p}\,\Theta(E - H(\mathbf{q},\mathbf{p};\lambda))\,,
\end{equation}
where $H$ is the system Hamiltonian and $\Theta(x)$ is the step function. For systems with one degree of freedom, $\Omega(E,\lambda)$ is an adiabatic invariant since it is the action \cite{Goldstein2002}. Thus, after a quasistatic change of $\lambda$, an initial energy shell is mapped into a final energy shell such $\Omega(E_{i},\lambda_{i}) = \Omega(E_{f},\lambda_{f})$. This relation can be seen as an expression for the final energy, $E^{ad}_{f}$, as function of $E_{i}$, $\lambda_{i}$ and $\lambda_{f}$ and hence gives the energy at the end of the quasistatic process.

\section{Excess work within linear response theory}\label{Appendix_A}

In this appendix we derive the linear response expression for excess and quasistatic work. In the inclusive picture, the thermodynamic work produced during a finite-time driving of a control parameter $\lambda$ is given by (\ref{total_work}). Assuming that $|\delta\lambda g(t)/\lambda_{0}|\ll 1$ for $t_{0} \leq t \leq t_{f}$, linear response theory provides the following relation between the out-of-equilibrium average and its corresponding response function $\phi_{0}(t)$ \cite{Kubo1}
\begin{eqnarray}
\overline{\dfrac{\partial \mathcal{H}}{\partial \lambda}}(t) = \left\langle \dfrac{\partial \mathcal{H}}{\partial \lambda} \right\rangle_{0} 
+ \chi_{0}^{\infty} \delta\lambda g(t) - \delta\lambda \int_{t_{0}}^{t} ds \, \phi_{0}(t-s) \, g(s) , \nonumber \\
\label{exc2}
\end{eqnarray}
where $\langle \cdot \rangle_{0}$ denotes the average on the initial microcanonical ensemble and the subscript refers to the value $\lambda_{0}$. The second term in the right-hand side of Eq.~(\ref{exc2}) describes the \textit{instantaneous} response, which is due to $\partial \mathcal{H}/\partial\lambda$ being a function of the external control $\lambda$. In particular, we have 

\begin{equation}
\chi_{0}^{\infty} = \left\langle \frac{\partial^{2} \mathcal{H}}{\partial\lambda^{2}} \right\rangle_{0}\ .
\end{equation}

The second term describes the \textit{delayed} response. It is convenient to express it in terms of the \textit{relaxation} function as $\phi_{0}(t) = - d\Psi_{0}(t)/dt$. Thus, integrating by parts Eq.~(\ref{exc2}) we find 
\begin{eqnarray}
\overline{\dfrac{\partial \mathcal{H}}{\partial \lambda}}(t) = \left\langle \dfrac{\partial \mathcal{H}}{\partial \lambda} \right\rangle_{0} &-& \delta\lambda \tilde{\Psi}_{0}(0) g(t) \nonumber \\
&+& \delta \lambda \int_{0}^{t-t_{0}} du \, \Psi_{0}(u) \dfrac{dg}{dt'}\bigg|_{t'=t-u}\,
\label{exc3}
\end{eqnarray}
where $\tilde{\Psi}_{0} \equiv \Psi_{0}(0) - \chi_{0}^{\infty}$. Finally, substituting Eq.~(\ref{exc3}) in expression (\ref{total_work}), we obtain 
\begin{eqnarray} \label{total_work2}
W &=& \delta\lambda \left\langle\dfrac{\partial \mathcal{H}}{\partial\lambda} \right\rangle_{0} - \dfrac{(\delta \lambda)^{2}}{2} \tilde{\Psi}_{0}(0) \nonumber \\
&+&  (\delta \lambda)^{2}  \int_{t_{0}}^{t_{f}} dt \, \frac{dg}{dt} \int_{t_{0}}^{t} dt' \, \Psi_{0}(t-t') \frac{dg}{dt'}\,,
\end{eqnarray}
where the following boundary conditions, $g(t_{0})=0$ and $g(t_{f})=1$, were used. The first two terms of the previous expression \textit{do not} depend on the protocol $g(t)$. Indeed, it can be verified that they are the first terms of the series expansion of the quasistatic work for $\delta\lambda/\lambda_{0} \ll 1$ given in (\ref{qs_work}). The last term clearly depends on $g(t)$ and therefore represents the \textit{excess} work given by (\ref{excess_work}). 
%

%%%%%%%%%%%%%%%%%%%%%%%%%%%%%%%%%%%%%%%%%%%%%%%%%%%%%%%%%%

%\bibliography{paper_MD2}

%

\end{document}